\documentclass[a4paper]{article}
\usepackage[latin1]{inputenc} 
\usepackage[T1]{fontenc} 
\usepackage{RR,RRthemes}
\usepackage{hyperref}
\usepackage{algorithm}
\usepackage{algorithmic}

\RRdate{Mars 2010}

\RRauthor{
Anya Apavatjrut\thanks[INSA]{CITI laboratory, INSA de Lyon, France}%
\and
Claire Goursaud\thanksref{INSA}
\and
Katia Jaffr\`{e}s-Runser\thanks[WNET]{WNET Laboratory, Stevens Institute of Technology, New Jersey, USA}\thanksref{INSA}
\and
 Cristina Comaniciu\thanksref{WNET}
\and
Jean-Marie Gorce\thanksref{INSA}
}
\authorhead{Apavatjrut\& Goursaud\& Jaffr\`{e}s-Runser\& Comaniciu\& Gorce}
\RRtitle{Stratégies de relayage pour exploiter de la diversité dans les réseaux de capteurs sans fils}
\RRetitle{Towards increasing diversity for the relaying of LT Fountain Codes in
Wireless Sensor Networks}
\titlehead{Relaying of LT Fountain Codes in
Wireless Sensor Networks}

\RRresume{La diversité est un outil important pour atteindre une bonne performance de transmission dans 
les réseaux sans fils. Dans cet article, nous proposons d'exploiter la diversité dans un réseau 
de capteurs à l'aide des différentes techniques de relayage en utilisant les codes fountaines. 
Contrairement à une transmission traditionnelle, chaque n\oe{}ud capteur participe
 à gérer et traiter les informations à transmettre. Nos algorithmes de relayage proposés 
sont basés sur l'opération XOR entre les paquets codés avec le code LT et nous avons montré que 
ces algorithmes donnent un gain de performance optimal en gardant la complexité du calcul
 basse pour convenir aux ressources des n\oe{}uds capteurs qui sont limitées.
}
\RRabstract{Diversity is a powerful means to increase the transmission performance of wireless communications.
For the case of fountain codes relaying, it has been shown previously that introducing diversity is also beneficial since it counteracts transmission losses on the channel. Instead of simply hop-by-hop forwarding information, each sensor node diversifies the information flow using XOR combinations of stored packets. This approach has been shown to be efficient for random linear fountain codes. However, random linear codes exhibit high decoding complexity.
In this paper, we propose diversity increased relaying strategies for the more realistic Luby Transform code in order to maintain high transmission performance with low decoding computational complexity in a linear network. Results are provided herein for a linear network assuming uniform imperfect channel states.
}
\RRmotcle{diversit\'e, algorithmes de relayage, relayage, code fontaine, LT, r\'eseau des capteurs}
\RRkeyword{diversity, relaying algorithms, fountain code, LT code, 
wireless sensor network, linear networks}
\RRprojets{SWING}
\RRdomaine{1} 
\RRthemeProj{SWING} 
\RRdomaineProjBis{SWING} 
\RCGrenoble 

\begin{document}
\RRNo{7231}
\makeRR   
\section{Introduction}
In a wireless sensor network (WSN), diversity has been shown to
significantly reduce the effect of fading, outage or nodes' failure. Diversity
can be exploited in time, frequency and space \cite{alamouti_1998}. Recently a
novel technique which consists in constantly retransmitting a XOR combination of the buffered packets as soon as a time slot becomes available was proposed in \cite{wicaksana_2008}. This method is based on a Random Linear fountain code (RL code). Performance gain can be achieved thanks to the increase in redundancy and diversity.

In a WSN, however, the use of RL code is limited due to its high computational complexity in the decoding process, especially when decoding is required at a node with limited
resources. Therefore, in this paper, we propose to study relaying
schemes with a more practical code called Luby Transform code (LT code). In contrast to a RL code, the decoding process of a LT code is more light-weight and is better suited for WSN. Indeed, its performance has been shown to approach Shannon capacity in a relay channel \cite{castura_2007} and it provides mutual information accumulation for multi-relay cooperative
transmissions \cite{molisch_2007}. Besides, as Raptor code is obtained by concatenating a precode with the LT code \cite{mackay_fountain_2005}, we just have to focus on the LT performance. 

Relaying with the LT code has been once mentioned in \cite{Gummadi_2008}. With this technique, each relay acts as an independent fountain encoder. The packets received by a relay are considered as raw data even though they have already been encoded several times. The decoder needs to peal off the successive encoded layers. As a matter of fact, this technique is too complex to be implemented on a WSN. 

The idea proposed in \cite{wicaksana_2008} where relay nodes simply combine packets using XOR operations is more attractive for WSN.
Although LT code itself has shown several benefits for WSN \cite{molisch_2007}, relaying schemes using an arbitrary XOR operation between LT encoded packets is subject to several constraints. Indeed, simply applying a XOR operation as in \cite{wicaksana_2008} becomes inefficient because the global degree distribution of the information flow arriving at the destination is altered and leads to decoding inefficiency. In this paper, we investigate several relaying solutions and we analyze their benefits and drawbacks. We highlight the tradeoff existing between the potential information throughput and the decoding efficiency. 

\section{Fountain codes}
Fountain codes are types of erasure codes that differ from other codes by their rateless property.  A source transmits an endless sequence of
encoded information to a destination while the decoding process ends as soon as the destination
has received enough packets.
This amount is usually slightly superior to the number of initial information due to additional redundancy.
The rate of the code is variable and adapts to the error ratio of each specific channel.
Another interesting feature of fountain codes is that they require limited feedback to achieve reliable transmission on unreliable links.
In this paper, we consider LT codes proposed in
\cite{luby_lt_2002} whose encoding and decoding processes can be described as
folllows.

\subsection{Encoding algorithm}
When information is to be sent from a source to a destination, it is first
partitioned into K fragments. These fragments have equal length and are encoded
together with a XOR operation to create a new sequence of encoded packets. The
number of fragments contained in each encoded packet is called "degree" of the
packet.
Luby described this "degree" as a key parameter to determine the coding
performance. A good degree distribution not only introduces sufficient redundancy
in the
information flow, but also facilitates the decoding procedure. Luby proposed the
"Robust Soliton Distribution" as an optimized degree distribution
\cite{luby_lt_2002}. This
degree distribution is designed so that the
decoding algorithm called Belief Propagation can be pursuit continuously and
efficiently.

\subsection{Decoding algorithm}
The decoding process is triggered as soon as the destination receives enough
packets
to be decoded. The most efficient decoding algorithm for any random codes
on an erasure channel is Maximum Likelihood decoding (ML-decoding). ML-decoding is
equivalent to solving systems of linear equations and can be performed using Gaussian
elimination. Although ML-decoding provides small error probability of decoding,
its complexity can grow rapidly with both N (number of received packets) and K
(code length), in the order of $\mathcal{O}(NK)$.

An alternative decoding technique which requires less computational
complexity is to use a graph based decoding method called Belief Propagation (BP-decoding).
BP-decoding solves linear equations based on a bipartite graph. At
each step, the decoder identifies an encoded packet that has degree one. If none
exists and the decoding process is not completed yet, the decoding process fails.
Otherwise, the
value of the original fragment contained in the identified packet is recovered
and the combination of this fragment will be removed from the rest of the
encoded
packets waiting to be decoded. These steps are repeated until the last fragment
is recovered successfully. BP-decoding algorithm has complexity of $\mathcal{O}(Kln(K))$.

\section{Relaying strategies}
In this work, we address the problem of designing an efficient multihop relaying strategy for a linear network where the source transmits a flow of packets encoded with an LT code.
The transmission is time-multiplexed and each received packet is relayed to the relay's immediate neighbor
on the next available time slot. An acknowledgement packet is sent by the destination once all information has been recovered.

The relaying algorithms proposed herein are compared to the two basic following benchmark strategies:
\paragraph*{\bf Strategy 1: Hop-by-hop Decode and Forward\bf }
Relay nodes totally decode and re-encode the packets before forwarding. This technique optimizes the number of transmitted packets by the source but introduces a high end-to-end delay as well as high decoding complexity at each relay.

\paragraph*{\bf Strategy 2: Passive Relaying\bf }
Relay nodes simply forward received packets once, without decoding.
There is no coding cost but transmission errors trigger a retransmission from the source which causes energy inefficiency.

In the last strategy, when a relay does not receive a packet, it simply waits for the next packet. The following strategies are designed to take advantage of the available time slot to introduce diversity in the information flow. Therefore, we add an internal buffer to each relay which stores the $B$ lastly received packets. When no packet is received in a time slot, the relay sends a combination of the packets from its buffer. The following strategies propose different algorithms designed to improve the transmission robustness for LT-codes.

\paragraph*{\bf Strategy 3: Last Packet\bf } The last received packet is retransmitted.
\paragraph*{\bf Strategy 4: XOR combination of R last packets\bf}
A XOR combination between the R last received packets is transmitted. This strategy is similar to the one
proposed for RL code in \cite{wicaksana_2008}. The main difference is that they send a XOR
combination every time slot while we only send one when a node has no packet to relay.

However, consecutive XOR combinations tend to increase the degree of each transmitted packet. Thus, BP-decoding usually fails due to the lack of degree one packets. To solve this issue, the following strategies concentrate on preserving the degree distribution of the LT code at the destination.

\paragraph*{\bf Strategy 5: XOR combination with prescribed degree\bf }
A XOR combination of the packets in the buffer is performed following Algorithm \ref{alg1}. In this algorithm, the degree $d$ of the output packet is chosen with
respect to the Robust Soliton Distribution. Buffered packets are then randomly selected and XOR-ed
together until degree $d$ is obtained or the $MAX\_ROUND$ value is reached.
\begin{algorithm}
\caption{XOR with prescribed degree}
\label{alg1}
\begin{algorithmic}
\STATE $p \Leftarrow $last received packet in the buffer
\STATE $d \Leftarrow $selected degree from Robust Soliton Distribution
\STATE $i \Leftarrow 0$
\WHILE{($i < MAX\_ROUND$)}
\STATE $prand \Leftarrow $randomly chosen packet from the buffer $\neq p$
\STATE $pxor \Leftarrow$ $p$ XOR $prand$
\IF{(degree of $pxor$ is closer to $d$ than degree of $p$)}
\STATE $p \Leftarrow$ $pxor$
\IF{(degree of $pxor$ = $d$)}
\STATE Return $pxor$
\ENDIF
\ENDIF
\STATE $i \Leftarrow$ $i$+1
\ENDWHILE
\end{algorithmic}
\end{algorithm}

\paragraph*{\bf Strategy 6: LT-Adapted XOR combination with prescribed degree\bf }
In practice, achieving low degree using Algorithm \ref{alg1} is quite difficult
for a
finite length buffer. Algorithm \ref{alg2} proposes an additional
improvement
that favours the transmission of low degree packets (e.g. degree 1 and 2) without XOR-ing. For an available time slot, if the last received packet has a low degree, it is either retransmitted with probability ($1-P_{XOR}$) or combined following Algorithm \ref{alg1}.

\begin{algorithm}
\caption{LT-Adapted XOR with prescribed degree}
\label{alg2}
\begin{algorithmic}
\STATE $p \Leftarrow $last received packet in the buffer
\STATE $P_{XOR} \Leftarrow $probability of XOR-ing
\IF{(degree of $p$ = 1 or 2)}
\IF{(rand[0,1]$\leq 1-P_{XOR}$)}
\STATE Retransmit $p$
\ELSE \STATE Return result from Algorithm~\ref{alg1}
\ENDIF
\ELSE
\STATE Return result from Algorithm~\ref{alg1}
\ENDIF
\end{algorithmic}
\end{algorithm}

\section{Simulation results}

\begin{figure}[!t]
\centering
\includegraphics[width=3in]{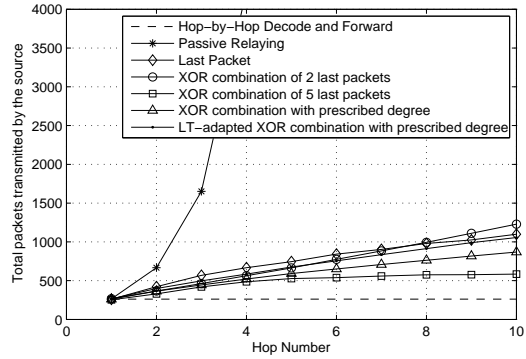}
\caption{Number of packets transmitted by the source with ML-decoding as a function of the hop number, with PER = 0.6, buffer size $B=100$, $MAX\_ROUND = 100$}
\label{fig:results_linear}
\end{figure}

\begin{figure}[!t]
\centering
\includegraphics[width=3in]{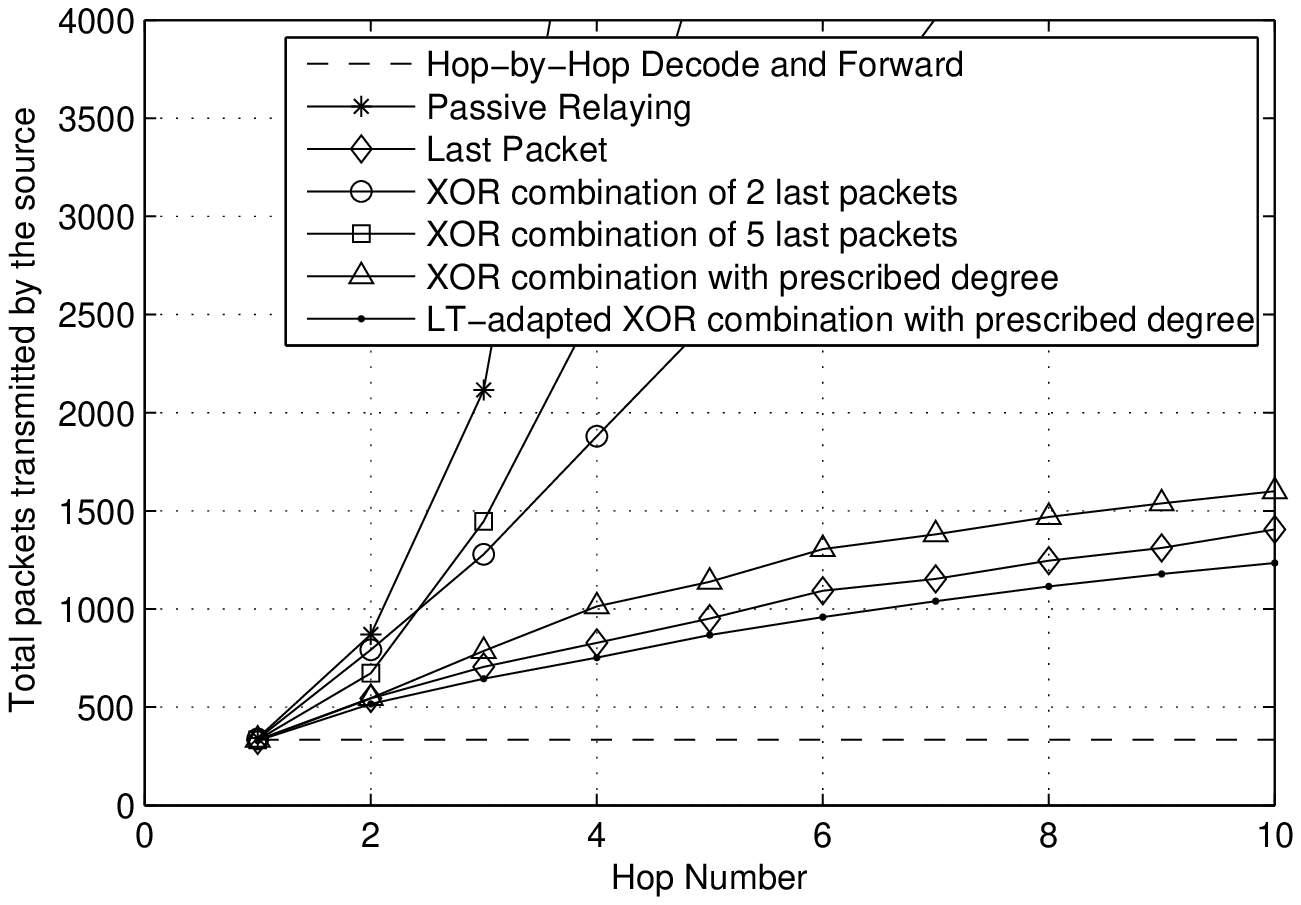}
\caption{Number of packets transmitted by source with BP-decoding as a function of the hop number, with PER = 0.6, buffer size $B=100$, $MAX\_ROUND = 100$}
\label{fig:results_BP}
\end{figure}

For any link of the linear network, we consider an erasure channel model characterized by
the same Packet Error Rate (PER). The performance gain of the relaying strategies is evaluated by the
number of packets transmitted by the source before decoding all the fragments at the destination. In our framework, it characterizes both
energy consumption and end-to-end delay. An extra header of size K containing
encoding information is added to each packet. The indices corresponding to the fragments of the original packet that are contained in the encoded packet are set to 1 while others are set to zero. Consequently, the header of a XOR-ed packet results from a XOR combination between the headers of the original packets. Figures \ref{fig:results_linear} and \ref{fig:results_BP} represent the number of transmitted packets as a function of the number of hops, for the ML and BP decoding algorithm respectively. The \textit{Hop-by-hop Decode and Forward} strategy gives the lowest bound on the number of sent packets. We have considered a channel with high PER to highlight the positive impact of our relaying strategies. 

Diversity performance is represented on Fig.~\ref{fig:results_linear}. This figure gives the total amount of packets needed to decode the information flow using an ML-decoder, i.e. the minimum bound of any decoding process for each relaying strategy. We notice that the relaying with retransmission strategies in an available time slot highly reduces the number of packets needed to be sent from the source. This confirms that XOR-ing along the line increases diversity and transmission efficiency. The more packets are combined together, the higher the robustness becomes. 

\begin{figure}[!t]
\centering
\includegraphics[width=3in]{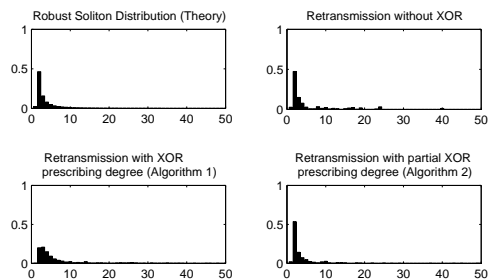}
\caption{Degree distribution for various relaying scenarios K=100, $\delta=0.5$, C=0.03}
\label{fig:results_distribution}
\end{figure}

This behavior is however different for the case of BP-decoding as shown in Fig.~\ref{fig:results_BP}. For the {\it XOR combination of 5 last packets} scenario, transmission diversity is at its best on Fig.~\ref{fig:results_linear} but for the BP-decoder, the severe alteration of the degree distribution limits its ability to decode. This feature is highlighted by Fig.~\ref{fig:results_distribution} which gives the degree distribution of the received packets after 10 hops.

For the BP-decoder, the basic strategy where the \textit{Last Packet} is retransmitted performs better than the retransmission of a \textit{XOR combination with prescribed degree}. As we can see from Fig.~\ref{fig:results_distribution}, low degree packets are very hard to be obtained from Algorithm \ref{alg1}. In contrast, the \textit{LT-Adapted XOR combination with prescribed degree} method
preserves the Robust Soliton Distribution and leads to efficient decoding process with high performance gain. From our simulations, the optimal probability $P_{XOR}$ is equal to 0.2.
With BP-decoding as shown in Fig.~\ref{fig:results_BP}, our proposed relaying technique outperforms all other existing relaying schemes because it trades off the resulting diversity and the decoding capability.

\section{Conclusion}
In this paper, we have proposed new relaying strategies for Fountain codes
that exploit diversity in a wireless sensor network. Our work is based on LT
code which is a pratical code that exhibits low complexity of decoding. We
investigate various relaying schemes that tradeoff diversity potential for
decoding capabilities. We have proposed a novel pratical scheme based on
probabilistic XOR-ing which preserves the Robust Soliton Distribution necessary
for Belief Propagation Decoding. Our simulation
results confirm that this relaying method outperforms all other relaying schemes.

\bibliographystyle{plain}
\bibliography{reference}
\end{document}